\documentclass[aps,prl,showpacs,floatfix,floats,nobalancelastpage,twocolumn]{revtex4-1}

\usepackage{graphicx}
\usepackage{graphics}
\usepackage{latexsym,amsmath,amssymb,bm,euscript}
\usepackage{color}
\usepackage{dcolumn}
\usepackage{bm, mathtools}
\usepackage{epstopdf}
\usepackage{times}
\usepackage{multirow}
\usepackage{wasysym}
\usepackage{subfigure}
\usepackage{bbm}
\usepackage{ulem}

\begin{document}

\title{Violation of Entanglement-Area Law in Bosonic Systems with Bose surfaces: Possible Application to Bose Metals}

\author{Hsin-Hua Lai}
\affiliation{National High Magnetic Field Laboratory, Florida State University, Tallahassee, Florida 32310, USA}

\author{Kun Yang}
\affiliation{Department of Physics and National High Magnetic Field Laboratory, Florida State University, Tallahassee, Florida 32306, USA}

\author{N. E. Bonesteel}
\affiliation{Department of Physics and National High Magnetic Field Laboratory, Florida State University, Tallahassee, Florida 32306, USA}

\date{\today}

\begin{abstract}
We show the violation of the entanglement-area law for bosonic systems with Bose surfaces. For bosonic systems with gapless factorized energy dispersions on a $N^d$ Cartesian lattice in $d$-dimension, e.g., the exciton Bose liquid in two dimension, we explicitly show that a belt subsystem with width $L$ preserving translational symmetry along $d-1$ Cartesian axes has leading entanglement entropy $(N^{d-1}/3)\ln L$. Using this result, the strong subadditivity inequality, and lattice symmetries, we bound the entanglement entropy of a rectangular subsystem from below and above showing a logarithmic violation of the area law. For subsystems with a single flat boundary we also bound the entanglement entropy from below showing a logarithmic violation, and argue that the entanglement entropy of subsystems with arbitrary smooth boundaries are similarly bounded. 
\end{abstract}
\maketitle

\textit{Introduction}---Entanglement is perhaps one of the most counter-intuitive aspects of quantum mechanics, and provides the sharpest distinction between quantum and classical descriptions of nature. It has been playing a growingly important role in characterization of phases and phase transitions in condensed matter physics. The most widely used measure of entanglement is the entanglement entropy (EE), which is the von Neumann entropy associated with the reduced density matrix of a subsystem, obtained by tracing out degrees of freedom outside it. For extended quantum systems, it is generally believed that ground states of all gapped local Hamiltonians, as well as a large number of gapless systems, follow the so called area law, which states that the EE is proportional to the surface area of the subsystem \cite{Eisert_RMP}. 

Violations of the area law, usually in a logarithmic fashion, do exist in various systems. They are found to be associated with quantum criticality in many one dimensional (1D) systems \cite{Calabrese2004}. However such violations are very rare above 1D; the {\em only} well-established examples in higher dimensions are free fermion ground states with Fermi surfaces, where it is found that the area law is enhanced by a logarithmic factor \cite{Wolf,GioevKlich,Swingle}. Recently, this result has been extended to Fermi liquid phases, and it was shown that Fermi liquid interactions do {\em not} alter the leading scaling behavior of the EE \cite{dingprx12}. Besides, it has also been used as a diagnostic of the presence of Fermi surface(s), even for non-Fermi liquid phases \cite{Huijse12, Zhang11}.

In contrast to fermionic systems, thus far there are no known quantum critical (or gapless) free bosonic systems that violate the area law above 1D \cite{Eisert_RMP, Barthel2006, Cramer2006}. The fundamental difference lies in the fact that gapless excitations normally live near a single point (usually the origin of momentum space) in such bosonic systems, while in Fermi liquids they live around an (extended) Fermi surface. In this work we show through explicit examples that logarithmic violation of the area law is possible in purely bosonic models above 1D. The models we use are motivated by the following considerations. Traditionally it was believed that bosons either condense (and become a superfluid) or localize (and insulate) at $T=0$. Recently it has been argued that, under certain circumstances, they can form so-called Bose metals with ``Bose surfaces,'' along which gapless excitations live \cite{Paramekanti02, LeeandLee05,ringxch, DBL,Sheng08, Sheng09,Tay10,Tay11,Mishmash11,Chua11,Biswas11,Baskaran09,Lai11}. These Bose surfaces resemble Fermi surfaces in intriguing ways, and may lead to violation of the area law. In this Letter we examine the EE of the so-called exciton Bose liquid (EBL) \cite{Paramekanti02, Tay10, Tay11} and show that it does indeed lead to such a violation.  The low-energy theory of the EBL is that of free bosons with an energy dispersion which vanishes linearly on a locus of points in $k$-space.  In our view this model plays the same ``idealized" role for Bose surface systems as the free fermion model does for Fermi surface systems \cite{bose_vs_fermi}.


Motivated by the long-wavelength description of the EBL in 2D, we study similar bosonic phases with gapless factorized energy dispersions in Cartesian systems in $d$D. We find that a belt subsystem $\mathcal{I}$ preserving translational symmetry along $d-1$ Cartesian axes explicitly shows a logarithmic violation of the area law, $S^d_{\mathcal I}\simeq(N^{d-1}/3) \ln L$, with $N$ and $L$ being the edge length of the whole Cartesian system and the width of the belt subsystem, respectively. Using lattice symmetries along with the strong sub-additivity inequality \cite{ArakiLieb1970, LiebRuskai1973, Lieb1973, Wehl1978,LiebRuskaiPRL,NielsenPetz2004,Casini2004, RyuTakayanagi2006, Headrick2007, Hirata2007, Fursaev2008, Hubeny2008}, we then find a lower bound on the EE of subsystems with a single flat boundary which also shows a logarithmic violation of the area law and argue that the EE of subsystems with arbitrarily smooth boundaries are similarly bounded. 

\textit{Bosonic Systems with Bose Surfaces}---The Lagrangian density for the long-wavelength bosonic theory of the EBL is \cite{Paramekanti02, Tay10, Tay11}
\begin{eqnarray}
{\mathcal L}_{EBL} = (\partial_\tau \vartheta)^2 + \kappa (\partial_x \partial_y \vartheta)^2,\label{EBL_lagrangian}
\end{eqnarray}
where $\kappa$ is the EBL ``phase stiffness" \cite{Tay11}, $\vartheta$ can be identified as the coarse-grain field dual to the boson phase $\phi$ in the bosonization of the 2D ring exchange model in Ref.~\cite{Paramekanti02}, and $\delta n = \pi^{-1} \partial_x \partial_y \vartheta$ is the parton-density fluctuation. 
The energy dispersion for the bosons, $\omega^2 \sim |k_x k_y|^2$, vanishes linearly on the $k_x$ and $k_y$ axes which together form a Bose surface.  

A lattice realization of this theory is provided by a 2D bosonic harmonic oscillator system on an $N\times N$ square lattice with factorized energy dispersion
\begin{eqnarray}\label{factorized_dispersion:2D}
\omega^2_{2D} = \left| f_x(k_x) f_y(k_y) \right|^2,~
\end{eqnarray}
where $f_x(k_x)$ and $f_y(k_y)$ are periodic functions of $k_x$ and $k_y$ which each vanishing linearly as $k_x$ and $k_y$ $\rightarrow 0$.  [The simplest example is $f_x(k_x) =2 \sin(k_x/2)$ and $f_y(k_y) =2 \sin(k_y/2)$ with lattice constant $a\equiv 1$.]

The oscillator Hamiltonian has the form
\begin{eqnarray}\label{HMO:H}
H=\frac{1}{2}\sum_{j} p_j^2 + \frac{1}{2}\sum_{j,k} q_j V_{jk} q_k,
\end{eqnarray}
where $q_j$ and $p_j$ are the displacement and momentum of oscillator $j$, respectively, and the elements of the matrix $V$ are determined by the inverse Fourier transform of the square of the energy dispersion (\ref{factorized_dispersion:2D}).  In the models studied here $V$ always describes short-ranged oscillator coupling.  Translational symmetry implies $V$ is a Toeplitz matrix; i.e., its elements depend only on the displacement $\vec r$ between oscillators $j$ and $k$, $V_{jk}\equiv V_{\vec{r}} = V_{-\vec{r}}$.  The factorized dispersion $\omega_{2D}^2$ further implies that $V$ iteslf is factorized with $V_{jk} = V^x_{x_j-x_k} V^y_{y_j-y_k}$ where the $V^x$ ($V^y$) matrix depends only on the $x$- ($y$-) component of the displacement between oscillators.  

Standard techniques can, at least in principle, be used to find the EE of a given subsystem in the ground state of the Hamiltonian (\ref{HMO:H}) \cite{Plenio2005, Unanyan2005, Cramer2006, Unanyan2007}. As pointed out in Ref.~\cite{Unanyan2005}, the particular factorized form of $V$ given above can give rise to a violation of the area law if $V^x$ and $V^y$ are both interaction matrices for 1D gapless harmonic  chains (as is the case for the EBL dispersion).  One technical issue is that for the EE to be well defined the matrix $V$ must be positive definite, with no zero eigenvalues.  The zero modes on the Bose surface must therefore be regularized.  One natural way to do this is to apply antiperiodic boundary conditions to the $N \times N$ lattice of oscillators. Doing so regularizes the zero modes without introducing a new length scale (other than the system size).  If we adopt this approach, the matrix $V$ satisfies the condition $V_{\vec{r}=(x,y)} = -V_{(x+N,y)}=-V_{(x,y+N)}=V_{(x+N,y+N)}$ and, for the dispersion (\ref{factorized_dispersion:2D}), the lowest eigenvalue of $V$ is of order $1/N^{2}$ \cite{regularization}. 

\begin{figure}[t] 
   \centering
   \includegraphics[width=1.1in]{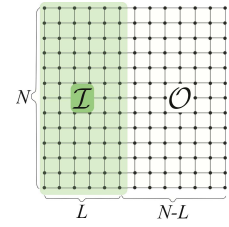} 
   \caption{(Color online) Partition of an $N \times N$ square lattice into the inner partition ($\mathcal{I}$) with size $L\times N$ and the outer partition ($\mathcal{O}$) with size $(N-L)\times N$. The subsystem $\mathcal{I}$ which preserves translational symmetry along one Cartesian axis is called the belt subsystem. The leading EE of the belt subsystem is shown in the text to be $S_{\mathcal I}=\frac{N}{3} \ln L$.}
 \label{Fig:Partition_HMO}
\end{figure}

Consider the EE of a block of oscillators with two flat edges separated by distance $L$ and parallel to a particular Cartesian axis (e.g., the $y$ axis, see Fig.~\ref{Fig:Partition_HMO}).  We refer to such a block as a belt subsystem and denote it ${\mathcal I}$ (the complementary subsystem of oscillators outside the block is denoted ${\mathcal O}$).  Following a procedure introduced by Cramer et al. \cite{cramer2007statistics} we perform a partial Fourier transform along the $y$-axis, $
\tilde q_{x,k_y} = \frac{1}{\sqrt{N}}\sum_y e^{ik_y y} q_{x,y}$, $\tilde p_{x,k_y} = \frac{1}{\sqrt{N}}\sum_y e^{ik_y y} p_{x,y}$. This transformation does not mix degrees of freedom in $\mathcal{I}$ with those in $\mathcal{O}$ and thus leaves the EE of the belt subsystem unchanged.  After this transformation the Hamiltonian is 
\begin{eqnarray}
\hspace{-0.3cm} H &=& \sum_{k_y} \frac{1}{2} \Bigg{[}\sum_x \tilde p_{x,k_y} \tilde p_{x,-k_y}\nonumber\\
&&~~~~~~~~~~ + f_y(k_y)\sum_{x_1,x_2} \tilde q_{x_1,k_y}  V^x_{x_1,x_2} \tilde q_{x_2,-k_y}\Bigg{]},
\label{transformed_H}
\end{eqnarray}
where (for antiperiodic boundary conditions) $k_y = (2 n_y+1)\pi/N$, with $n_y=0, 1, ..., N-1$.  

The Hamiltonian (\ref{transformed_H}) describes $N$ decoupled 1D chains with dispersions $\omega_{1D}(k_x) = |f_x(k_x)| |f_y(k_y)|$ with $k_y$ fixed and nonzero.  Each chain then has $\omega_{1D}(k_x) \sim |k_x|$ for $k_x \ll 1$ and so is conformally invariant at long wavelengths. We therefore expect all $N$ chains to contribute $S_{1D} = \frac{1}{3} \ln L$ to the EE in the $L \gg 1$ limit \cite{Holzhey1994, Keating2004, Korepin2004, Calabrese2004}.   If the zero modes are regularized using antiperiodic boundary conditions then the matrix $V^x_{x_k - x_j}$ is an antiperiodic function of $x_k-x_j$ with antiperiod $N$ [For the case $f_x(k_x) = 2 \sin (k_x/2)$,   $V^x(0) = 2, V^x(1) = -1, V^x(n) = 0$ for $n\in[2,N-1]$ and $V^x(N) = +1$ corresponding to oscillators coupled by nearest-neighbor springs].  For this regularization scheme, if $L$ is held fixed and $N$ taken to $\infty$ the EE will diverge (as seen explicitly in the numerical work of Ref.~\cite{skrovseth2005entanglement}).  We therefore consider the limit $N,L \rightarrow \infty$ keeping the ratio $L/N$ fixed.  In this limit we indeed expect the leading contribution to the EE of each chain to be $S_{1D} \simeq \frac{1}{3} \ln L$ for large enough $L$ regardless of the $L/N$ ratio.  Numerical studies of the gapless harmonic 1D chain with nearest neighbor coupling and antiperiodic boundary conditions confirm this expectation \cite{skrovseth2005entanglement}. 

The $N$ 1D chains described by Hamiltonian (\ref{transformed_H}) are critical.  They each violate the 1D area law logarithmically and contribute $\frac{1}{3} \ln L$ to the EE of the belt subsystem $\mathcal{I}$.  The total EE of ${\mathcal I}$ is thus  
\begin{eqnarray}\label{EE:TI_2D} 
S_{\mathcal I}~\dot{=}~\frac{N}{3} \ln L,
\end{eqnarray}
which violates the 2D area law (here $\dot{=}$ indicates the leading contribution in the scaling limit described above). This violation occurs because the number of critical chains contributing to $S_{\mathcal I}$ is extensive in $N$. This in turn is a direct consequence of the existence of a Bose surface and should be contrasted with the case of critical Bose systems with dispersions vanishing at a single point in $k$ space considered in Ref.~\cite{cramer2007statistics} for which the number of critical chains is not extensive in $N$ and the area law holds. Note that the simplest case with $f_x(k_x) = 2 \sin (k_x/2)$ and $f_y(k_y) = 2 \sin (k_y/2)$ can be realized in Hamiltonian (\ref{HMO:H}) with only first and second neighbor couplings.

\begin{figure}[t]
\subfigure[]{\label{Fig:Partition_3D} \includegraphics[width=1.2in]{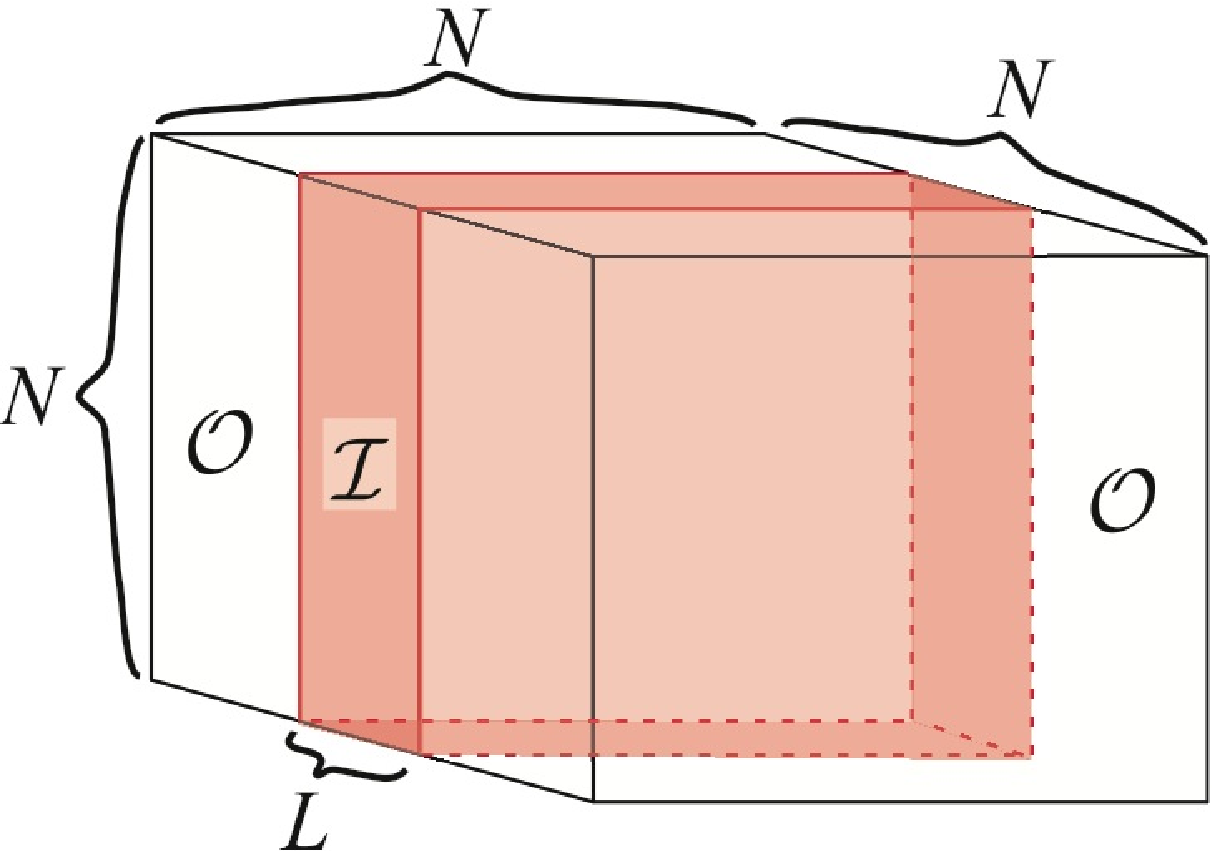}}
\subfigure[]{\label{Fig:Rectangle_2D}\includegraphics[width=1.0in]{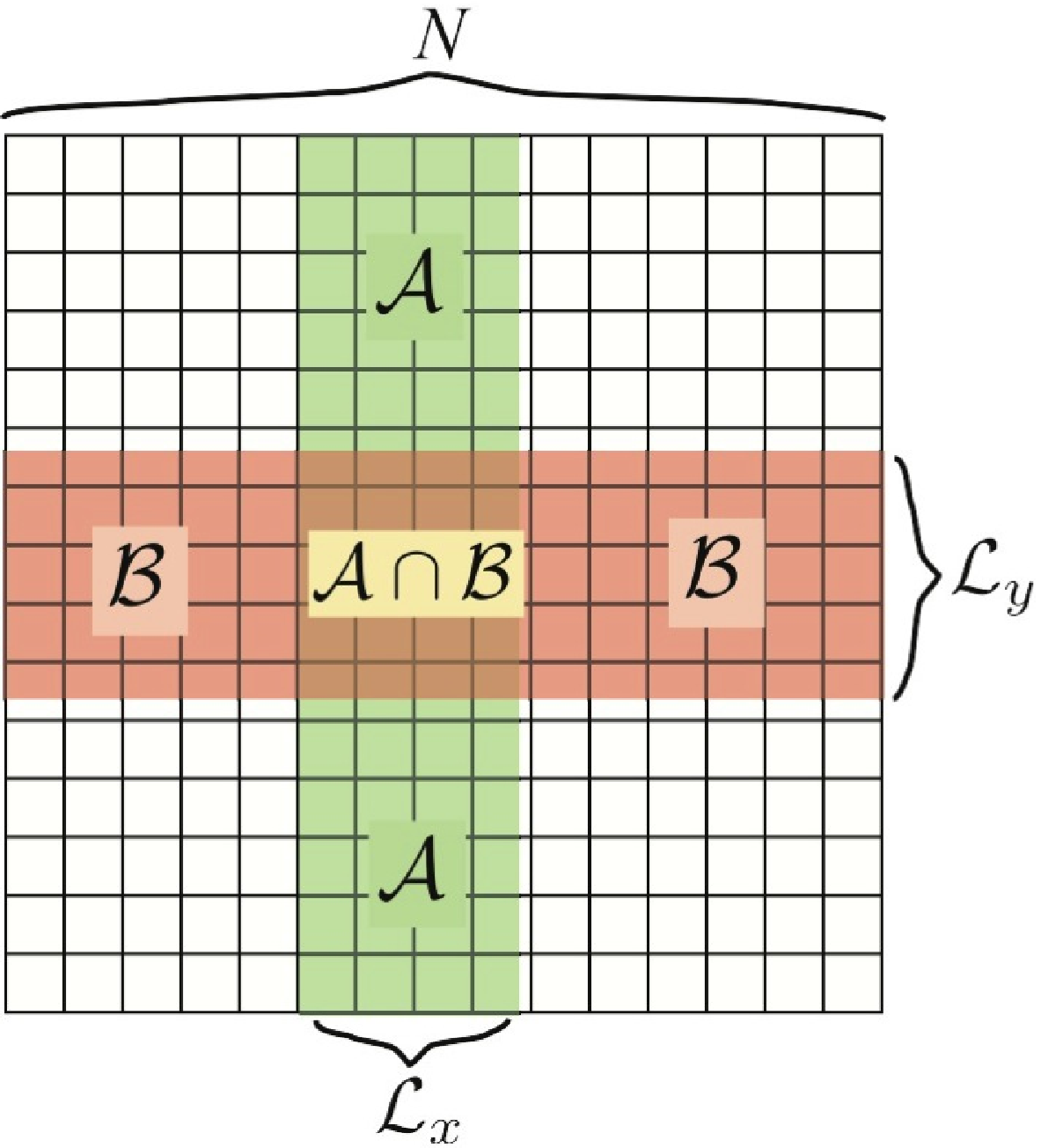}}
\subfigure[]{\label{Fig:Equal_Par_2D}\includegraphics[width=1.0in]{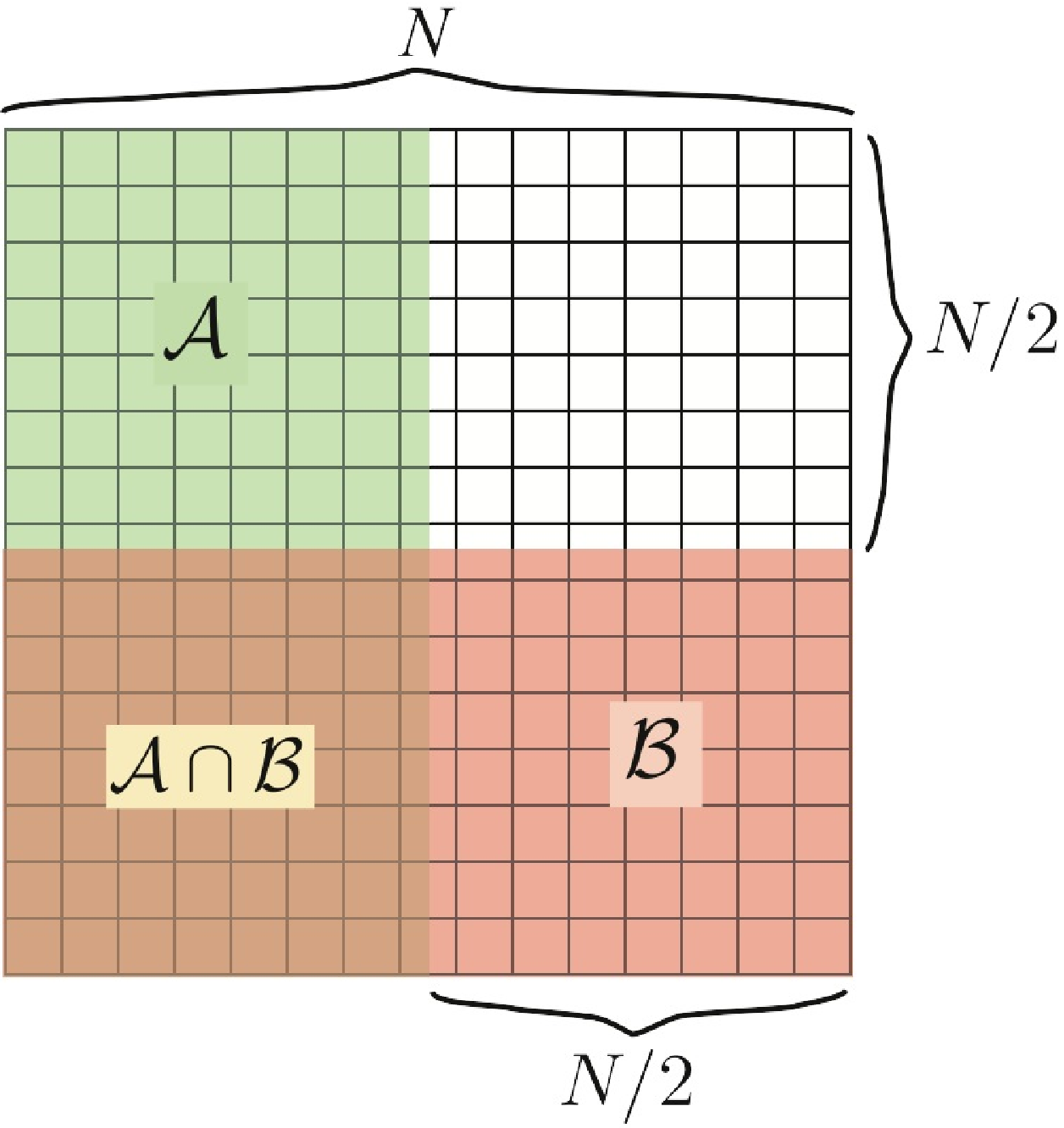}}
\caption{(Color online) (a) Partition of a 3D lattice into the subsystem $\mathcal{I}$ and $\mathcal{O}$. The shaded box regime $\mathcal{I}$ with size $L\times N\times N$ is the belt subsystem of interest. (b)  Rectangular subsystem of a 2D lattice which can be considered as the overlap region of two belt subsystems $\mathcal{A}$ with size $\mathcal{L}_x \times N$ and $\mathcal{B}$ with size $N\times \mathcal{L}_y$. The leading entanglement entropies of the belt subsystems are $S_{\mathcal{A}}=\frac{N}{3}ln \mathcal{L}_x$ and $S_{\mathcal{B}} = \frac{N}{3} \ln \mathcal{L}_y$. (c) Illustration of equal partition in 2D. Each subsystem has size $N/2 \times N/2$.}
\label{Fig:Partition_3D_2D}
\end{figure}

The result (\ref{EE:TI_2D}) can be generalized straightforwardly to belt subsystems in $d\geq 2$ [see, e.g., Fig.~\ref{Fig:Partition_3D} for the 3D version of the partition]. The EE in $d$D for a system with dispersion $\omega^2_{dD} = |f_1(k_1)f_2(k_2)f_3(k_3)\cdots f_d(k_d)|^2$ where each $f_i(k_i)$ vanishes linearly as $k_i \rightarrow 0$ is 
\begin{eqnarray}\label{EE:TI_general}
S^d_{\mathcal I}~\dot{=}~\frac{N^{d-1}}{3} \ln L.
\end{eqnarray}
 
\textit{Bounds for EE of rectangular subsystems in $d$D}---A rectangular subsystem on a 2D square lattice can be viewed as the intersection of two perpendicular belt subsystems $\mathcal{A}$ and $\mathcal{B}$, (see Fig.~\ref{Fig:Rectangle_2D}). The region $\mathcal{A}\cap\mathcal{B}$ has length $\mathcal{L}_x$ and width $\mathcal{L}_y$. We can put an upper bound on the EE of this rectangular subsystem using the strong subadditivity inequality  \cite{ArakiLieb1970, LiebRuskai1973, Lieb1973, Wehl1978,LiebRuskaiPRL,NielsenPetz2004,Casini2004, RyuTakayanagi2006, Headrick2007, Hirata2007, Fursaev2008, Hubeny2008},
\begin{eqnarray}\label{EE:rec_above}
\nonumber && S_{\mathcal{A}} + S_{\mathcal{B}} \geq S_{\mathcal{A} \cup \mathcal{B}} + S_{\mathcal{A} \cap \mathcal{B}} \geq S_{\mathcal{A}\cap \mathcal{B}}\\
&&\Rightarrow \frac{N}{3} \ln (\mathcal{L}_x \mathcal{L}_y) \geq S_{\mathcal{A} \cap \mathcal{B}}\equiv S_{\Box},
\end{eqnarray}
where we assume $N,~\mathcal{L}_x,~\mathcal{L}_y,~N-\mathcal{L}_x,~N-\mathcal{L}_y \gg 1$ \cite{LvsN}.  Here we explicitly use Eq.~(\ref{EE:TI_2D}) and the positivity of the EE. 
 
To obtain a lower bound, we consider $N= n_x \mathcal{L}_x$ and $N=n_y \mathcal{L}_y$, where $n_x,~n_y \in \mathbb{R}_+$. By translational symmetry, the EE of any subsystem with the same shape, size, and orientation as $\mathcal{A}\cap\mathcal{B}$ is equal to $S_{\Box}$. We can therefore clone $\lceil n_{y} \rceil$ ($\lceil n_{x} \rceil$) copies of $\mathcal{A} \cap \mathcal{B}$ and pile them along the $y$-direction ($x$-direction) to cover the whole area of subsystem $\mathcal{A}$ $(\mathcal{B})$, each of which has the same EE, $S_1=S_2=\cdots=S_{\lceil n_{y} \rceil } =S_{\Box}$ and a similar relation for $y \leftrightarrow x$. By the strong subadditivity inequality we have $ \nonumber  S_1 + S_2 + \cdots S_{\lceil n_y \rceil}  \geq S_{1\cup2\cup3\cup \cdots \cup\lceil n_y \rceil} = S_{\mathcal{A}} \Rightarrow S_{\Box} \geq \frac{\mathcal{L}_y}{3}  \ln \mathcal{L}_x$ and a similar relation with $y\leftrightarrow x$. The EE of a rectangular subsystem can then be bounded below as 
\begin{eqnarray}
\label{EE:rec_below}
S_{\Box}\geq max.\left[ \frac{\mathcal{L}_x}{3}\ln \mathcal{L}_y, \frac{\mathcal{L}_y}{3}\ln \mathcal{L}_x\right].
\end{eqnarray}

For a concrete example, let us consider a partition of the 2D system into four ($2^2$) equally-sized square subsystems. The belt subsystems $\mathcal{A}$ and $\mathcal{B}$ now are each half of the whole system (see Fig.~\ref{Fig:Equal_Par_2D}) and we are interested in placing upper- and lower- bounds on $S_{\mathcal{A}\cap \mathcal{B}}$. In this partitioning, $S_{\mathcal{A} \cup \mathcal{B}} = S_{complement}=S_{\mathcal{A}\cap\mathcal{B}}$, and we obtain a better upper bound than Eq.~(\ref{EE:rec_above}) because $S_{\mathcal{A}} + S_{\mathcal{B}} \geq S_{\mathcal{A}\cup\mathcal{B}} + S_{\mathcal{A}\cap \mathcal{B}} = 2 S_{\mathcal{A}\cap\mathcal{B}}$. The EE $S^{2}_{EP} \equiv S^{2, EP}_{\Box} $ (where EP indicates an equally partitioned region) is bounded as
\begin{eqnarray}
\frac{N}{3}\ln \left( \frac{N}{2}\right) \geq S^{2, EP}_{\Box} \geq \frac{N}{6} \ln \left( \frac{N}{2}\right).
\end{eqnarray}

For $d>2$, the above argument is straightforwardly generalized to show that the EE of the equally-partitioned subsystem ($2^d$ equally-sized subsystems) in $d$-dimension, $S^d_{EP}\equiv S^{d, EP}_\Box$ can be bounded as
\begin{eqnarray}
\frac{d N^{d-1}}{3}\ln \left(\frac{N}{2}\right) \geq S^{d, EP}_{\Box} \geq \frac{1}{3} \left( \frac{N}{2}\right)^{d-1}\ln \left(\frac{N}{2}\right).~~~
\end{eqnarray}

\begin{figure}[t]
\subfigure[]{\label{Fig:1edge_initial} \includegraphics[width=1in]{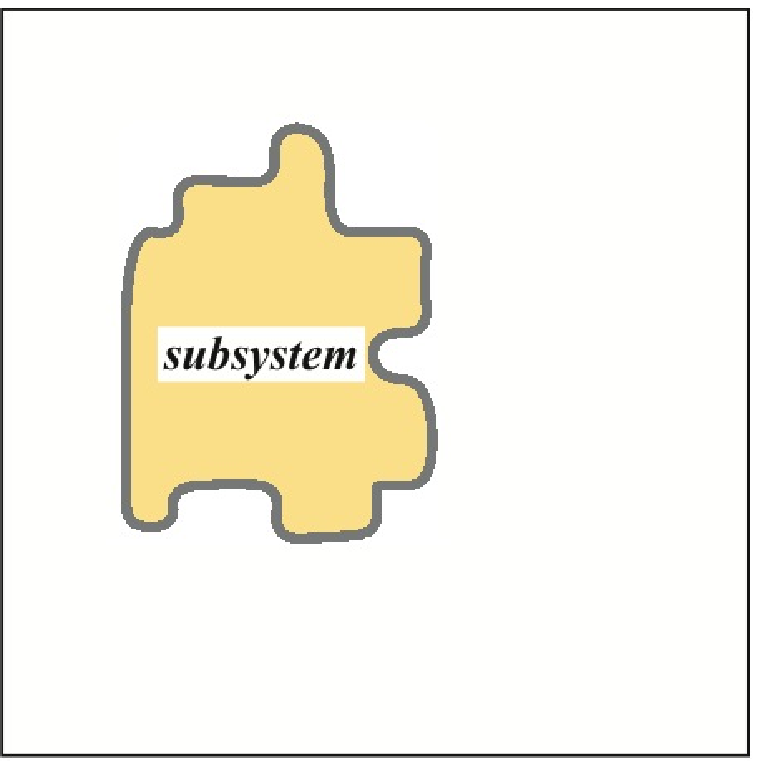}}\hspace{0.3cm}
\subfigure[]{\label{Fig:1edge_clone} \includegraphics[width=1in]{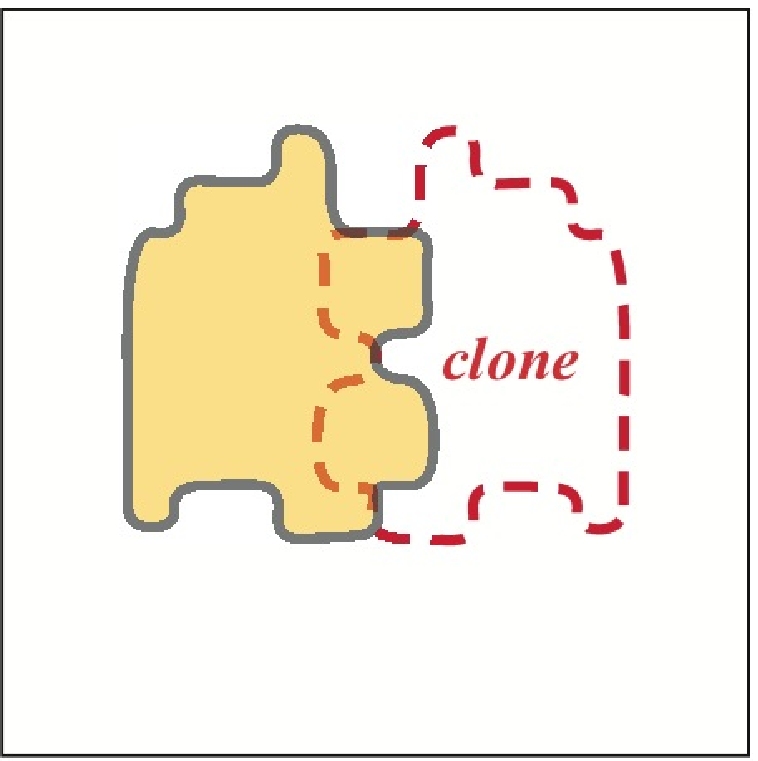}}\\
\subfigure[]{\label{Fig:1edge_tile} \includegraphics[width=1in]{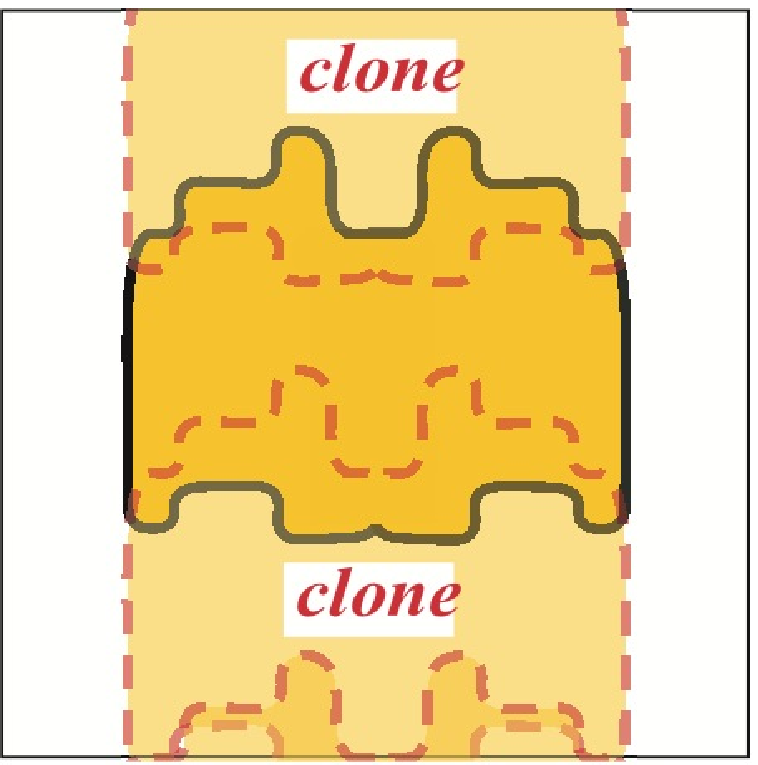}}\hspace{0.4cm}
\subfigure[]{\label{Fig:arbitrary_argument}\includegraphics[width=1in]{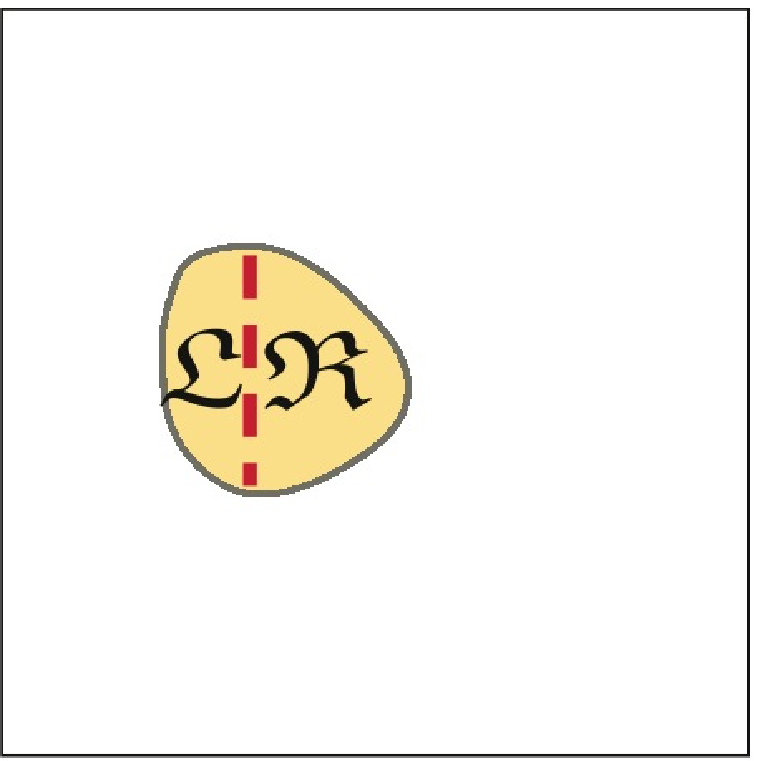}}
\caption{(Color online) (a)-(c) Illustrations of the use of mirror and lattice translation symmetries to form a belt system to obtain a logarithmic entropic lower bound.  The shaded region in (a) represents the initial subsystem with one flat boundary. In (b) we clone a copy of the subsystem and use mirror symmetry to flip it. We then overlap the flipped subsystem with the original one to make a new subsystem. In (c) we clone the new subsystem in (b) and tile it along the Cartesian axis (y-axis here) to form a belt system, which gives the logarithmic entropic lower bound.  (d) Arbitrary slicing along a Cartesian axis of a subsystem with general smooth boundary into a left region $\mathfrak{L}$ and a right region $\mathfrak{R}$.}
\label{Fig:arbitrary}
\end{figure}

Now consider a non-rectangular subsystem with at least one boundary parallel to a Cartesian axis, Fig.~\ref{Fig:1edge_initial}. We can still use lattice symmetries and the strong subadditivity inequality to obtain a lower bound on the EE. Taking the EBL as an explicit example, which has lattice translation and mirror symmetries, first we clone the original subsystem and use mirror symmetry to flip the cloned subsystem about a Cartesian axis. We then overlap the clone with the original to form a new subsystem with two parallel flat edges, see Fig.~\ref{Fig:1edge_clone}. Next, we make copies of the new subsystem and tile them along a Cartesian axis to form the belt subsystem, Fig.~\ref{Fig:1edge_tile}. Finally, focusing on the belt subsystem, we can adopt the result (\ref{EE:TI_2D}) to show a logarithmic lower bound on the EE. 

We have not found a rigorous way to establish a violation of the area law for subsystems with general smooth boundary. However, we consider it likely that such a violation does occur. A general subsystem can be ``arbitrarily" sliced into a left region ($\mathfrak{L}$) and a right region ($\mathfrak{R}$) by a cut parallel to a Cartesian axis as shown in Fig.~\ref{Fig:arbitrary_argument}.  Since $\mathfrak{L}$ and $\mathfrak{R}$ both have a flat boundary parallel to a Cartesian axis, the arguments given above imply that $S_\mathfrak{L}$ and $S_\mathfrak{R}$ will both show logarithmic enhancements to the area law.  The only way that the EE of the full subsystem $S_{\mathfrak{L} \cup \mathfrak{R}}$ would not have a similar enhancement would be if the leading logarithmic enhancements of $S_\mathfrak{L}$ and $S_\mathfrak{R}$ were each entirely due to entanglement of oscillators in $\mathfrak{L}$ with those in $\mathfrak{R}$.  Given that the division of the general subsystem into $\mathfrak{L}$ and $\mathfrak{R}$ is arbitrary we view such a cancellation as implausible.  Rather, we believe the logarithmic enhancement is due to long range correlations in the system.  We thus expect that a subsystem with general smooth boundary shows a logarithmic violation of the entropic area law. Such arguments can be straightforwardly generalized to $d>2$. 

\textit{Arbitrary Bose Surface}---The arguments presented here are not unique to the factorized EBL dispersion.  We expect similar logarithmic enhancement of the area law for systems with generic Bose surfaces.  For example, a system of harmonic oscillators with dispersion
\begin{eqnarray}
\omega^2_{\bf k} = \alpha^2\left[ \sin^2 (k_x/2) + \sin^2 (k_y/2) - \beta\right]^2,\label{closedFS}
\end{eqnarray}
has a closed Bose surface for $0 < \beta < 2$. To compute the EE of a belt subsystem ${\cal I}$ of width $L$ one can again follow Ref.~\cite{cramer2007statistics} and Fourier transform along the direction parallel to the boundary of ${\cal I}$, say the $y$-direction,  to obtain decoupled chains with dispersions given by Eq.~(\ref{closedFS}) for fixed values of $k_y$.  

\begin{figure}[t] 
\subfigure[]{\label{} \includegraphics[width=1.1in]{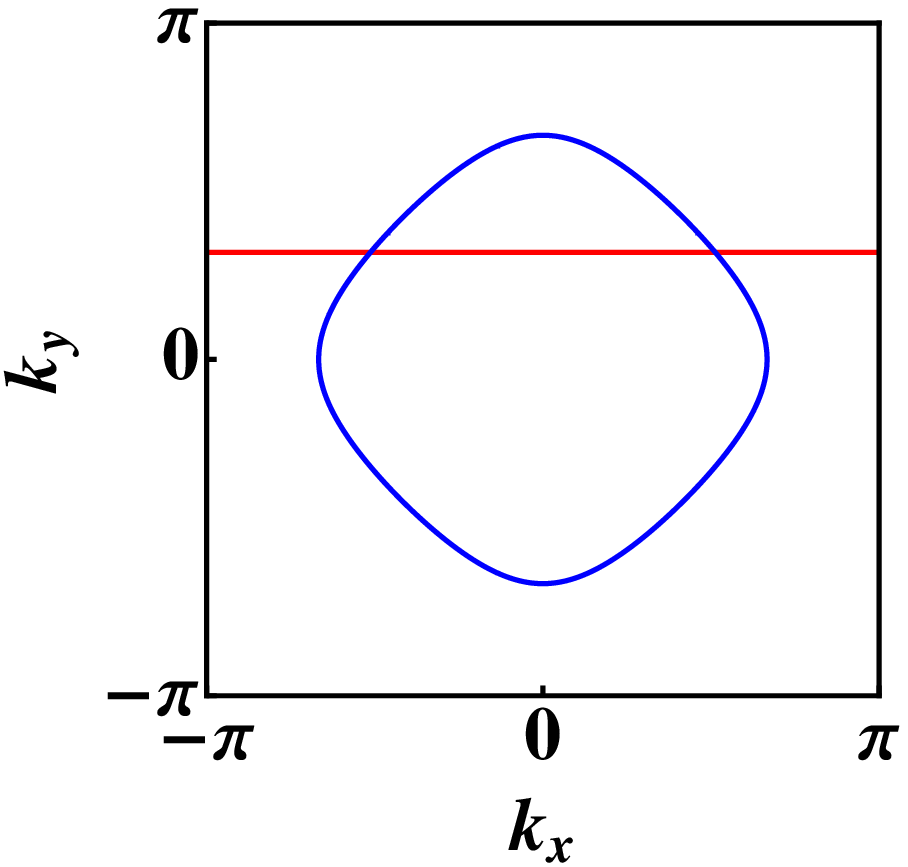}}\hspace{0.3cm}
\subfigure[]{\label{}\includegraphics[width=1.3in]{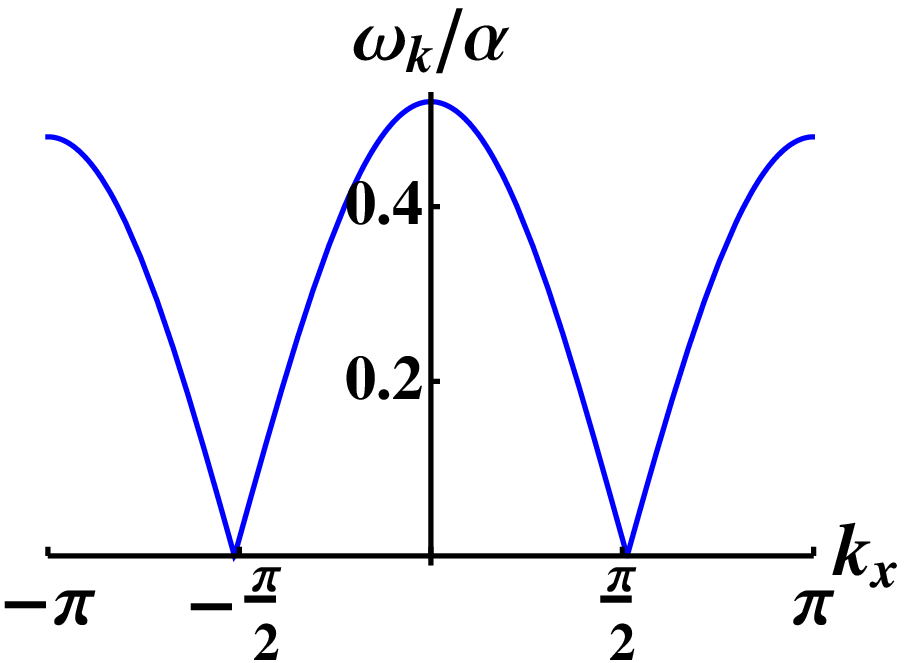}}
   \caption{(Color Online) (a) Closed Bose surface for the  dispersion (\ref{closedFS}) with $\alpha=1$ and $\beta=0.75$ and a line of constant $k_y$ corresponding to one of the decoupled chains which contributes to the entanglement entropy of a belt subsystem ${\cal I}$ with boundary parallel to the $y$-axis. (b) 1D dispersion of the decoupled chain corresponding to the line in (a).  }
   \label{Fig:closed}
\end{figure}

As illustrated in Fig.~\ref{Fig:closed}, for those values of $k_y$ which correspond to lines that intersect the Bose surface, the resulting 1D dispersion is critical and generically has two gapless points where the dispersion vanishes linearly.  We therefore expect each such chain to contribute $\frac{2}{3} \ln L$ to the EE of ${\mathcal I}$ which will then be $S_{\mathcal I} \simeq \gamma N \ln L$ where $\gamma$ is a geometric factor associated with the size of the Fermi surface along the $k_y$ axis. The strong subadditivity arguments given above then imply a logarithmic enhancement to the area law for rectangular and more general subsystems.  As also noted above, the essential ingredient for this enhancement is the extended Bose surface which gives rise to an extensive number of critical chains contributing to $S_{\mathcal I}$ after a partial Fourier transform.  

\textit{Conclusions}---In this work we show that the entanglement entropy of Bose metals has a logarithmic violation of the area law. We explicitly study bosonic systems with gapless factorized energy dispersions, such as the exciton Bose liquid in 2D. We explicitly give the entanglement entropy of the belt subsystems in $d-$dimension which shows logarithmic enhancement. We bound the entanglement entropy of the subsystems with at least a single flat boundary in a way that shows the logarithmic violation and argue that subsystems with arbitrarily smooth boundary are similarly bounded. The implication of this work is that entropic area law violation is perhaps more common than thought. It is not a unique identifier of the presence of Fermi surface in fermionic systems, as it can also be associated with Bose metals. 

HHL and KY acknowledge the support from the National Science Foundation through Grant No. DMR-1004545. NEB acknowledges support from US DOE Grant No. DE-FG02-97ER45639.

\bibliography{biblio4EE}

\begin{thebibliography}{46}
\bibitem [{Lvs()}]{LvsN}%
  \BibitemOpen
  \href@noop {} {}\bibinfo {note} {In the regime considered here, with $N,
  {\mathcal L}_{x}, {\mathcal L}_y \gg 1$ and fixed ratios $N/{\mathcal L}_{x}$
  and $N/{\mathcal L}_y$, the logarithmic factor enhancing the area law can be
  expressed as either $\ln{\mathcal L_x}$, $\ln{\mathcal L_y}$, or $\ln N$.
  Here we choose to use $\ln{\mathcal L}_{x}$ and $\ln{\mathcal L}_y$ to
  clarify where the bound came from and the relevant Cartesian
  axis.}\BibitemShut {Stop}%
\end{thebibliography}

\end{document}